\documentclass[twocolumn,showpacs,superscriptaddress,amsmath,amssymb]{revtex4}
\usepackage{graphicx}
\usepackage{caption2}
\usepackage{dcolumn}
\usepackage{bm}
\newcommand\figcaption{\def\@captype{figure}\caption}

\def\e{{\rm e}}  \def\d{{\rm d}}

\def\ie{{\it i.e. }}   \def\cf{{\it cf. }}
  
 \def\fc{\frac}  
            
    \def\cA{{\cal A}}  \def\sgn{{\rm sgn}}
\def\bcc{\begin{center}} \def\ecc{\end{center}}
\def\beq{\begin{equation}} \def\eeq{\end{equation}}
\def\bea{\begin{eqnarray}} \def\eea{\end{eqnarray}}
\def\beqa{\begin{eqnarray}}  \def\eeqa{\end{eqnarray}}
  
\def\btbl{\begin{tabular}} \def\etbl{\end{tabular}}
\def\bpm{\begin{pmatrix}} \def\epm{\end{pmatrix}}
\def\btbb{\begin{tabbing}} \def\etbb{\end{tabbing}}
\def\btm{\begin{itemize}} \def\etm{\end{itemize}}
  
  \def\nbr{\nonumber}
\def\f{\left}  \def\g{\right}  
\def\ched{\end{CJK*} \end{document}} \def\bs{\boldsymbol}
     
\def\pt{p_{\rm t}}

\def\ack{{\vskip5mm\noindent\bf Acknowledgements\ }}
\def\bd{
\begin{document}}    \def\ed{\end{document}}
\def\bpm{\begin{pmatrix}} \def\epm{\end{pmatrix}}
\def\ben{\begin{enumerate}} \def\een{\end{enumerate}}
\def\btb{\begin{tabular}} \def\etb{\end{tabular}}
\def\btbb{\begin{tabbing}} \def\etbb{\end{tabbing}}
\def\qqg{q$\bar{\rm q}$g\hskip1pt} \def\EE{e$^+$e$^-$\hskip1pt }
\def\pt{{p_{\rm t}}} \def\bpt{\bar{p_{\rm t}}}
\def\NAtwo{{\sc na}{\footnotesize 22}}  \def\QGP{{\sc qgp}}
\def\RHIC{{\sc rhic}}  \def\BNL{{\sc bnl}}\def\ebe{event-by-event}
\def\nba{$N\hskip-2pt Ba$\hskip5pt}

\bd   

\title{The inversion-asymmetry of pion emission source along the\\
outward direction in relativistic heavy ion collisions}
\author{Liu Lianshou\footnote{Email:
liuls@iopp.ccnu.edu.cn} \ \   Shi Shusu\ and  \ Du Jiaxin}
\affiliation{Institute of Particle Physics, Huazhong Normal
University, Wuhan 430079, China}

\begin{abstract}
The inversion-asymmetry of the pion emission source in
relativistic heavy ion collision under the Bertsch-Pratt
convention is discussed and explicitly exhibited by a Monte Carlo
model. The Gaussian source function popularly used in the HBT
analysis of relativistic heavy ion collisions is invalid in this
case. An inversion-asymmetric source function is suggested. A
method for extracting the inversion-asymmetry degree of the source
together with the source size from experimental data is proposed.
\end{abstract}


\maketitle

The analysis of two-particle, in particular two-pion, correlation
provides a powerful tool for determining the spatio-temporal
characteristics of relativistic heavy ion
collisions\;\cite{review}. It is usually referred to as pion
interferometry or HBT analysis, following the pioneer work of
Hanbury-Brown and Twiss\;\cite{HBT}. Unfortunately, there is no
one-to-one correspondence between the two-particle correlation
function in momentum space and the source distribution in
configuration space. How to extract as much information as
possible about the source distribution from the experimentally
measured two-particle correlation data is the main challenge for
the HBT analysis.

Two-particle correlation function can be expressed as \beq C(\bs
P,\bs q)=1+ \int d^3 \bs r S_{\bs P}(\bs r) |\phi_{\bs q}(\bs
r)|^2 , \eeq where $\bs P ={\bs p_1+\bs p_2}$ is the total
momentum of the pair and $\bs q=\bs p_1-\bs p_2$ is their relative
momentum. $|\phi_{\bs q}(\bs r)|^2$ is a weight factor, which
takes care of the indistinguishability of identical particles.
$S_{\bs P}(\bs r)$ is the source function. The integral transform
in Eq.\;(1) does not have a unique inverse due to the ``on shell''
constraint $\bs P\cdot\bs q=0$\;\cite{uli}. Therefore, to extract
the distribution $S_{\bs P}(\bs r)$ from the experimentally
measurable correlation $C(\bs P,\bs q)$ additional assumption must
be made on the functional form of the source function $S_{\bs
P}(\bs r)$, which should be consistent with the symmetry of the
system.

Two kinds of symmetry are related to the present problem.  The
first kind is the rotational symmetry, \ie the symmetry property
with respect to the space rotation. For a central heavy ion
collision there is axial symmetry about the beam axis, defined as
{\it longitudinal direction}. This amounts to isotropic in the
transverse plane, \ie the source appears as a circle in the
transverse plane with the same radius in all directions. For a
non-central or peripheral collision, the axial symmetry is
violated. The source has an ellipse form in the transverse plane.
The radii along the two main axes are nonequal.

The second kind of symmetry is the inversion symmetry, \ie the
symmetry property with respect to the space inversion $x_i \to
-x_i$. This symmetry holds for all the three coordinates in the
laboratory frame.  However, in the present relativistic heavy ion
experiments the HBT analysis is carried out not in the laboratory
frame but in the Bertsch-Pratt frame\;\cite{Bertsch}\cite{Pratt}
--- $x_l, x_o, x_s$, where the {\it
longitudinal} component $x_l$ still points along the beam
direction; the {\it outward} component $x_o$ is defined as
parallel to the projection $\bs K$ of the pair momentum $\bs P$ on
the transverse plane; the {\it sideward} component $x_s$ is
perpendicular to both $x_l$ and $x_o$. This coordinate system has,
in particular, the advantage that the ratio $x_o/x_s$ is sensitive
to the duration of
hadronization\;\cite{Bertsch}\cite{Pratt}\cite{gyulassylecture}.
It is essentially a pair-by-pair rotation of the laboratory frame
so that the new coordinate $x_o$ points to the pair transverse
momentum $\bs K$ for all the pairs, \cf Fig.\;1.

In  the Bertsch-Pratt frame, the inversion transform of the
coordinate $x_o$ \beq x_o\rightarrow -x_o\eeq will change the
direction of the vector $\bs K$, and therefore, is no longer a
symmetry transform of the system. In other words, the existence of
a fixed vector $\bs K$ along the direction $x_o$ violates the
inversion symmetry about $x_o=0$, and a Gaussian source becomes
invalid along this direction.

\begin{figure}[!ht]
\begin{center}
\includegraphics[width=0.5\linewidth]{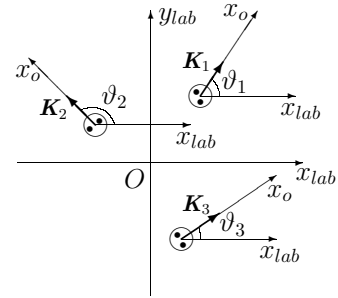}
\caption{\label{phn} The pair-by-pair rotation from laboratory
frame to Bertsch-Pratt frame. In the figure $\vartheta_i$ is the
rotation angle of $x_{\rm lab}$ for the $i$th pair.}
\end{center}
\end{figure}

The aim of the present letter is to discuss how to take the
inversion asymmetry along $x_o$ into account in constructing a
source function.

The inversion symmetry is an internal property of the
distribution, which is defined in the intrinsic frame where the
average value of coordinate vanishes,\beq
\int_{-\infty}^{\infty}x\, P(x)\d x=0 .\eeq Thus by definition a
Gaussian is always inversion symmetric.

The {\it inversion-asymmetry degree} $\cA$ can be defined in the
above-mentioned frame as
\def\rnba{P}
\beq \cA=\sqrt{\fc{\int_{-\infty}^\infty
\f[\rnba(x)-\rnba(-x)\g]^2\d x}{2\int_{-\infty}^\infty
[\rnba(x)]^2 \d x}}\ \ ,\eeq which vanishes for inversion
symmetric distribution and equals unity for maximum asymmetric
distribution, where $P(x) \neq 0$ only when $P(-x)=0$.

In  current relativistic heavy ion experiments while performing
HBT analysis a Gaussian source function is usually used. Such a
function may be isotropic or anisotropic consistent with the
rotational symmetric  or asymmetric of the collision, but could
never be inversion asymmetric.

As stated above, in relativistic heavy ion collisions the
inversion symmetry holds for the source distribution in the
laboratory frame, where the axes in the transverse plane are fixed
and are common for all the particle pairs, but is broken along the
$x_o$ direction of the Bertsch-Pratt frame, \cf Fig.\;1, so the
Gaussian source function is no longer valid along this direction.

\begin{figure}
\includegraphics[width=2.8in]{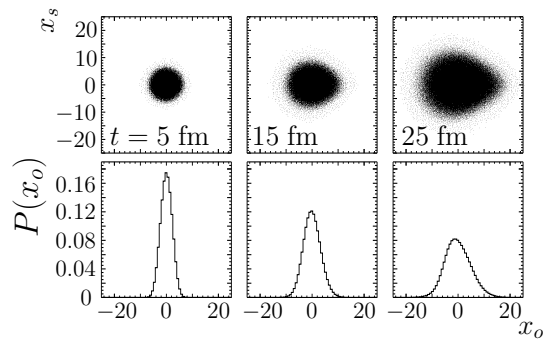}
\caption{\label{Fig. 1} Upper row: the $x_o$-$x_s$ distribution of
hadron pairs in a typical event of Au-Au central collision at
$\sqrt{s_{\rm NN}}=200$ GeV  from AMPT simulation at three
different time. From left to right $t=5, 15, 25$ fm, respectively.
The lower row is the projection to the $x_o$ axis.}\end{figure}

In order to illustrate how the $x_o$ distribution violates the
inversion symmetry and how the inversion-asymmetry develops in the
heavy ion collision process, we take a Monte Carlo generator
AMPT\;\cite{ampt} as example. This model is based on
non-equilibrium transport dynamics. It contains four main
components: the initial conditions, partonic interactions,
conversion from the partonic to the hadronic matter and hadronic
interactions. The initial conditions, which includes the spatial
and momentum distributions of minijet partons from hard processes
and strings from soft processes, are obtained from the HIJING
model\;\cite{hijing} in which eikonized parton model is employed.
The time evolution of partons is then treated according to the
ZPC~\cite{zpc} parton cascade model. After partons stop
interacting, a combined coalescence and string fragmentation model
is used for the hadronization of partons. Scattering among the
resulting hadrons are described by a relativistic transport (ART)
model~\cite{art} which includes baryon-baryon, baryon-meson and
meson-meson elastic and inelastic scattering.

In the upper row of Fig\;2 is sketched the $x_o$-$x_s$
distribution of hadron pairs in a typical event of $\sqrt{s_{\rm
NN}}=200$ GeV Au-Au central collision at three different times
--- $t=5, 15, 25$ fm. The inversion-asymmetry along $x_o$ can clearly be seen.
The asymmetry degree develops stronger and stronger with the
increase of time. The lower row is the projection to the $x_o$
axis.

In order to extract the asymmetry information of source we need a
source distribution function which can extrapolate from nearly
symmetric, nearly Gaussian, to highly asymmetric. For this purpose
we propose the following three-parameter $N$, $B$, $a$
distribution \beq P(x)=\begin{cases}
\fc{1}{B^{N+1}\Gamma(N+1)}(a\mp x)^N \e^{\fc{\pm x-a}{B}} & \pm
x<a , \cr 0 & \pm x>a.
\end{cases}\eeq

Transforming to the intrinsic frame, satisfying Eq.\;(3), we have
\begin{widetext}
\beq P_{N\hskip-2pt Ba}(x)=\begin{cases} \sgn(B)\cdot
\fc{[(N+1)B-x)]^N}{B^{N+1}\Gamma(N+1)} \;
 \e^{\fc{x-(N+1)B}{B}} & \mbox{\ \rm for\ \ }
 \sgn(B)\cdot x< \sgn(B)\cdot (N+1)B , \cr 0 &  \mbox{\ \rm for\ \ } \sgn(B)\cdot
 x>
 \sgn(B)\cdot (N+1)B,
\end{cases}\eeq
where \beq N>0,\qquad
 \sgn(B)=\begin{cases} +1
 & B>0 , \cr
 -1 & B<0.
\end{cases}
\eeq
\end{widetext}

In Fig.\;3 are plotted the $\rnba_{N\hskip-2pt Ba}(x)$'s for three
different $N$. It can be seen that the distribution varies from
approximately symmetric about $x_o=0$ at large $N$ to highly
asymmetric at small $N$. At sufficiently large $N$ the \nba
distribution Eq.\;(6) can mimic the Gaussian distribution very
well.
 \begin{figure}
 \includegraphics[width=3.2in]{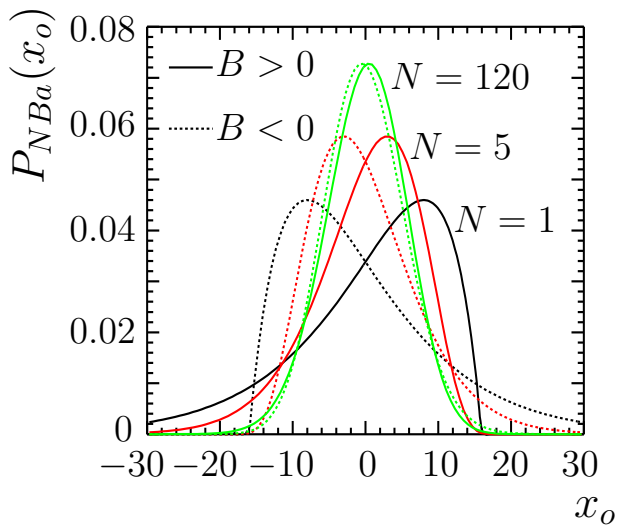}
 \caption{\label{Fig. 3} (Color on line) The \nba distributions for
 3 different values of $N$. The absolute values of $B$ are adjusted
 to make the 3 curves have successively lower height.}
 \end{figure}

It turns out that the asymmetry degree $\cA$ of this distribution
\beq \cA= \sqrt{ 1-\fc {\Gamma(\fc{1}{2})\Gamma(N+1)}
{\Gamma(2N+1)\Gamma(N+\fc{3}{2})}
\cdot[2(N+1)]^{2N+1}\e^{-2(N+1)}} \eeq
depends only on the parameter $N$ but not on $B$ and $a$. The
asymmetry degree $\cA$ as function of $N$ is plotted in Fig.\;4.
It can be seen that $\cA$ varies from $\sim 0$ at large $N$ to
$\sim 0.5$ at small $N$.

 \begin{figure}
 \includegraphics[width=2.6in]{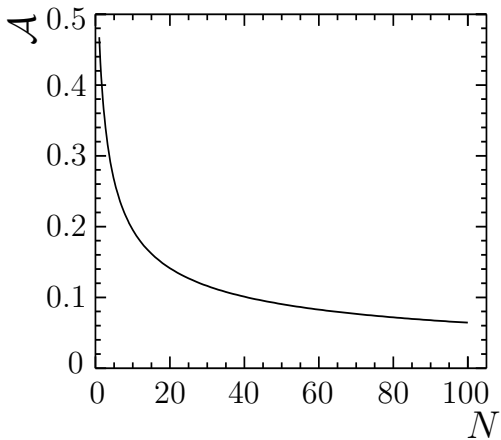}
 \caption{\label{Fig. 3} The asymmetry degree of \nba
 distribution.}
 \end{figure}

 \begin{figure}
 \includegraphics[width=3.2in]{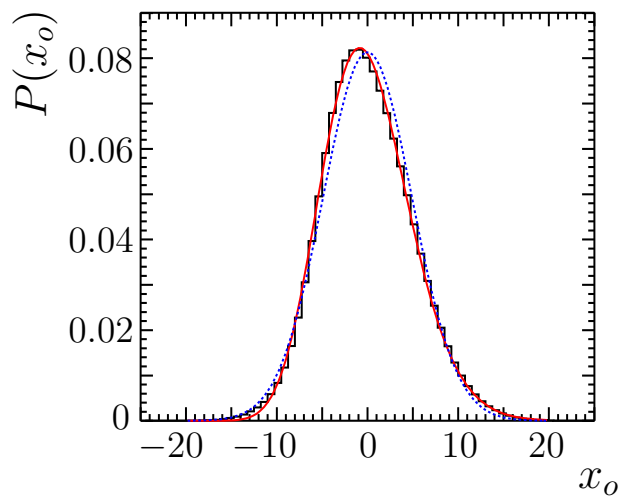}
 \caption{\label{Fig. 4} (Color on line) The fit of the $x_o$
 distribution of the AMPT event at $t=25$ fm used in Fig.\;2 to \nba
 (solid line) and Gaussian (dashed line).}
 \end{figure}

\begin{table}[h]\caption{\label{table1}   The size $R_o$ and
asymmetry degree $\cA$ of a typical $\sqrt{s_{\rm NN}}=200$ GeV
Au-Au central collision event at $t=5, 15, 25$ fm used in Fig.\;2,
obtained from the fit of the $x_o$ distribution to the \nba and
Gaussian distributions, respectively}
\begin{center}
\begin{tabular}{|c|c c|c c|}
\hline  & \multicolumn{4}{c|}{Distribution}      \\
  & \multicolumn{2}{c|}{\nba}  & \multicolumn{2}{c|}{Gaussian}   \\
\hline  Time   &  $R_o$ &  $\cA$ &  $R_o$ &  $\cA$    \\
\hline 5 fm   & 2.28   &  0.0065 &  2.28 &  ---   \\
\hline 15 fm  & 3.30   &  0.0565 &  3.30  &  ---   \\
\hline 25 fm  & 4.90   &  0.1137 &  4.90 &  ---  \\
\hline
\end{tabular}
\end{center}
\end{table}

The size, or ``radius'', $R$ defined as the root mean square of
coordinate $x$ is \beq R_{N\hskip-2pt
Ba}=\sqrt{\int_{-\infty}^{\infty} x^2 P_{N\hskip-2pt Ba} (x)\d x
}=|B|\cdot\sqrt{N+1}.\eeq

We fit the $x_o$ distributions of the Monte Carlo events shown in
the lower row of Fig.\;2 to the \nba distribution (6) and present
the results for $t=25$ fm in Fig.\;5 as solid line. For comparison
the same distribution is also fitted to Gaussian and shown in the
same figure as dashed line. It is evident that the \nba fit can
account for the inversion-asymmetry of $x_o$ distribution present
in the Monte Carlo data, while the Gaussian fit fails to do so.
The size $R_o$ and asymmetry degree $\cA$ resulting from the fit
are listed in Table\;I. Using the \nba source function we are able
to extract the asymmetry degree of the source, which could not be
achieved using a Gaussian source function. As for the source size
the results from both \nba and Gaussian source functions are the
same.

The inversion asymmetry exists only in the outward direction, so
we assume Gaussian sources in $x_l$ and $x_s$, while \nba source
in $x_o$, \beq P(\bs r) \sim \e^{-\f[ \fc{x_l^2}{2R_l^2}+
\fc{x_s^2}{2R_s^2}\g]} \cdot P_{N\hskip-2pt Ba}(x_o). \eeq
Performing Fourier transform we get \beq \rho(\bs q)\sim
\fc{\e^{-\fc{1}{2}(q^2_lR^2_l+q^2_sR^2_s)}\e^{-iq_o(N+1)B}}{(1-iq_o
B)^{N+1}}. \eeq The correlation function is \bea C_2(\bs
q)&=&1+\lambda |\rho(\bs q)|^2 \nbr\\ &=& 1+\lambda
\,\fc{1}{[1+q_o^2B^2]^{(N+1)}} \,\e^{-(q^2_lR^2_l+q^2_sR^2_s)}.
\eea

Through fitting the experimentally measured correlation function
to Eq.\;(12) the parameters $R_l$, $R_s$, $N$ and $B$ are
obtained. From these parameters beside being able to extract the
size parameters $R_l$ and $R_s$ in the longitudinal and sideward
directions, we are also armed with a powerful tool for extracting
the asymmetry degree $\cA$ and size parameter $R_o$ in the outward
direction by using Eq's.\;(8) and (9).

It is argued that under the Bertsch-Pratt
convention\;\cite{Bertsch}\cite{Pratt} popularly used in the HBT
analysis of relativistic heavy ion collision there is a fixed
vector
--- the total transverse momentum $\bs K$ of the pair, along the
outward direction.  The existence of such a vector violates the
inversion symmetry along this direction, making the conventional
Gaussian source function invalid. Although the size of source can
be extracted from the experimental data using a Gaussian source
function, an important feature of the system
--- the asymmetry degree along $x_o$ is missed.

In the present letter the inversion-asymmetry along $x_o$ is
pointed out in the first time. A 3-parameter source distribution
------ the \nba distribution is proposed, which can extrapolate
from nearly inversion-symmetric to highly inversion-asymmetric.
Using a source function, which is Gaussian in the longitudinal and
sideward directions, while \nba in the outward direction, the
asymmetry degree $\cA$ along the outward direction can be
extracted from the experimental data together with the source-size
parameters $R_l$, $R_s$ and $R_o$.

The inversion asymmetry along $x_o$ in a typical Au-Au central
collision event under the Bertsch-Pratt convention has been
checked using the AMPT Monte Carlo model, and the successful
extraction of the asymmetry degree $\cA$ from the data is
exhibited. To apply the proposed method to real experimental data
is highly recommended.

\vskip8mm \ack This work is supported in part by the National
Science Foundation of China under project 10375025 and by the
Cultivation Fund of the Key Scientific and Technical Innovation
Project, Ministry of Education of China NO CFKSTIP-704035. The
authors thank W. A. Zajc, M. A. Lisa and Wu Y. F. for helpful
discussions.

\end{document}